\documentclass[lettersize,journal]{IEEEtran}
\usepackage{amsmath,amsfonts}
\usepackage{algorithmic}
\usepackage{algorithm}
\usepackage{array}
\usepackage[caption=false,font=normalsize,labelfont=sf,textfont=sf]{subfig}
\usepackage{textcomp}
\usepackage{stfloats}
\usepackage{url}
\usepackage{verbatim}
\usepackage{graphicx}
\usepackage{cite}
\usepackage{xcolor}
\usepackage{tabularx}
\usepackage{array}
\usepackage{flushend}
\begin{document}
\addtolength{\textfloatsep}{-0.25in}
\title{Memory-Immersed Collaborative Digitization for Area-Efficient Compute-in-Memory Deep Learning}
\author{Shamma Nasrin$^1$, Maeesha Binte Hashem$^1$, Nastaran Darabi$^1$, Benjamin Parpillon$^1$, Farah Fahim$^2$, Wilfred Gomes$^3$, and Amit Ranjan Trivedi$^1$ \\
$^1$AEON Lab, University of Illinois at Chicago (UIC), Chicago, IL, USA, $^2$Fermi National Accelerator Lab (FNAL), Batavia, IL, USA, $^3$Intel Corp., Hillsboro, OR, USA}
\maketitle

\begin{abstract}
This work discusses memory-immersed collaborative digitization among compute-in-memory (CiM) arrays to minimize the area overheads of a conventional analog-to-digital converter (ADC) for deep learning inference. Thereby, using the proposed scheme, significantly more CiM arrays can be accommodated within limited footprint designs to improve parallelism and minimize external memory accesses. Under the digitization scheme, CiM arrays exploit their parasitic bit lines to form a within-memory capacitive digital-to-analog converter (DAC) that facilitates area-efficient successive approximation (SA) digitization. CiM arrays collaborate where a proximal array digitizes the analog-domain product-sums when an array computes the scalar product of input and weights. We discuss various networking configurations among CiM arrays where Flash, SA, and their hybrid digitization steps can be efficiently implemented using the proposed memory-immersed scheme. The results are demonstrated using a 65 nm CMOS test chip. Compared to a 40 nm-node 5-bit SAR ADC, our 65 nm design requires $\sim$25$\times$ less area and $\sim$1.4$\times$ less energy by leveraging in-memory computing structures. Compared to a 40 nm-node 5-bit Flash ADC, our design requires $\sim$51$\times$ less area and $\sim$13$\times$ less energy.   
\end{abstract}


\section{Introduction}
A memory structure stores model parameters and performs most inference computations for the compute-in-memory (CiM) processing of deep neural networks (DNN). Thus, by integrating model storage and computations, CiM averts significant data movements between intermediate memory hierarchy and processing modules that plague the performance of conventional digital architectures for DNN. Even more, traditional memory structures such as SRAM \cite{sehgal2021trends}, RRAM \cite{yu2021compute}, embedded-DRAM \cite{jung2022dualpim,xie202116}, \textit{etc.}, can be adapted for CiM, making the scheme highly attractive for cost-effective adoption in various systems-on-chip (SOC). Most CiM schemes also leverage analog representations of operands, such as element-wise weight-input product terms represented as charge or current, to simplify their summation over a wire by exploiting Kirchhoff's law. Thus, exploiting physics can minimize the necessary workload and processing elements. 

Although intermediate analog representations of operands provide critical advantages to CiM, such as improving parallelism, minimizing workload, and simplifying compute cells, they also present significant implementation challenges for CiM peripherals. For analog computations, a CiM array requires digital-to-analog converters (DAC) and analog-to-digital converters (ADC) to operate on digital inputs and digitize the analog output for routing and storage. The analog circuits are susceptible to failure under process variability, posing challenges for area and energy scalability. Recently, many techniques have obviated analog DACs from CiM, such as by encoding the inputs as time pulses \cite{hung20228}, employing bit slicing \cite{sakr2021signal}, using multiplication-free operators \cite{nasrin2021mf}, \textit{etc.} Meanwhile, the necessity of ADC in a CiM array is directly linked with utilizing charge or current-based representation of the product values to avoid a dedicated adder. Hence ADCs are still necessary for many schemes.

\begin{figure}[t!]
  \centering
  \includegraphics[width=0.85\linewidth]{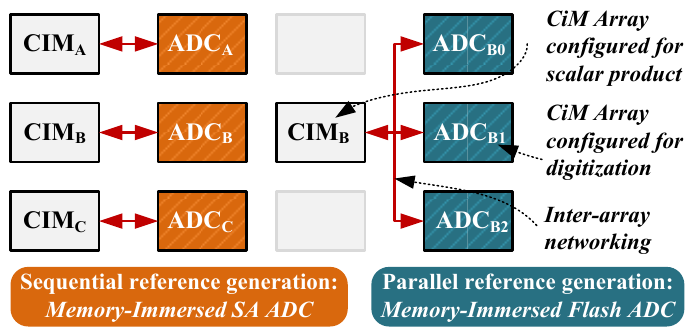}
  \caption{\textbf{Overview of memory-immersed collaborative digitization:} At the left, neighboring CiM arrays are coupled for sequential reference generation for memory-immersed successive approximation ADC. At the right, one CiM array is coupled with many arrays to the right for parallel reference generation for memory-immersed Flash ADC.}
  \label{fig:mainfig}
\end{figure}

This work discusses a novel \textit{memory-immersed digitization} that can preclude a dedicated ADC and its associated area overhead. Fig. \ref{fig:mainfig} shows a high-level overview of the scheme. Thereby, with simpler peripherals, most of the silicon space can be used only for CiM. In the proposed digitization scheme, parasitic bit-lines of memory arrays form within-memory capacitive DAC, and neighboring memory arrays collaborate for resource-efficient digitization. By configuring the networking among proximal CiM arrays flash, successive approximation (SA), and their hybrid operations can be performed efficiently within memory arrays. 

\begin{figure*}[t!]
  \centering
  \includegraphics[width=0.58\linewidth]{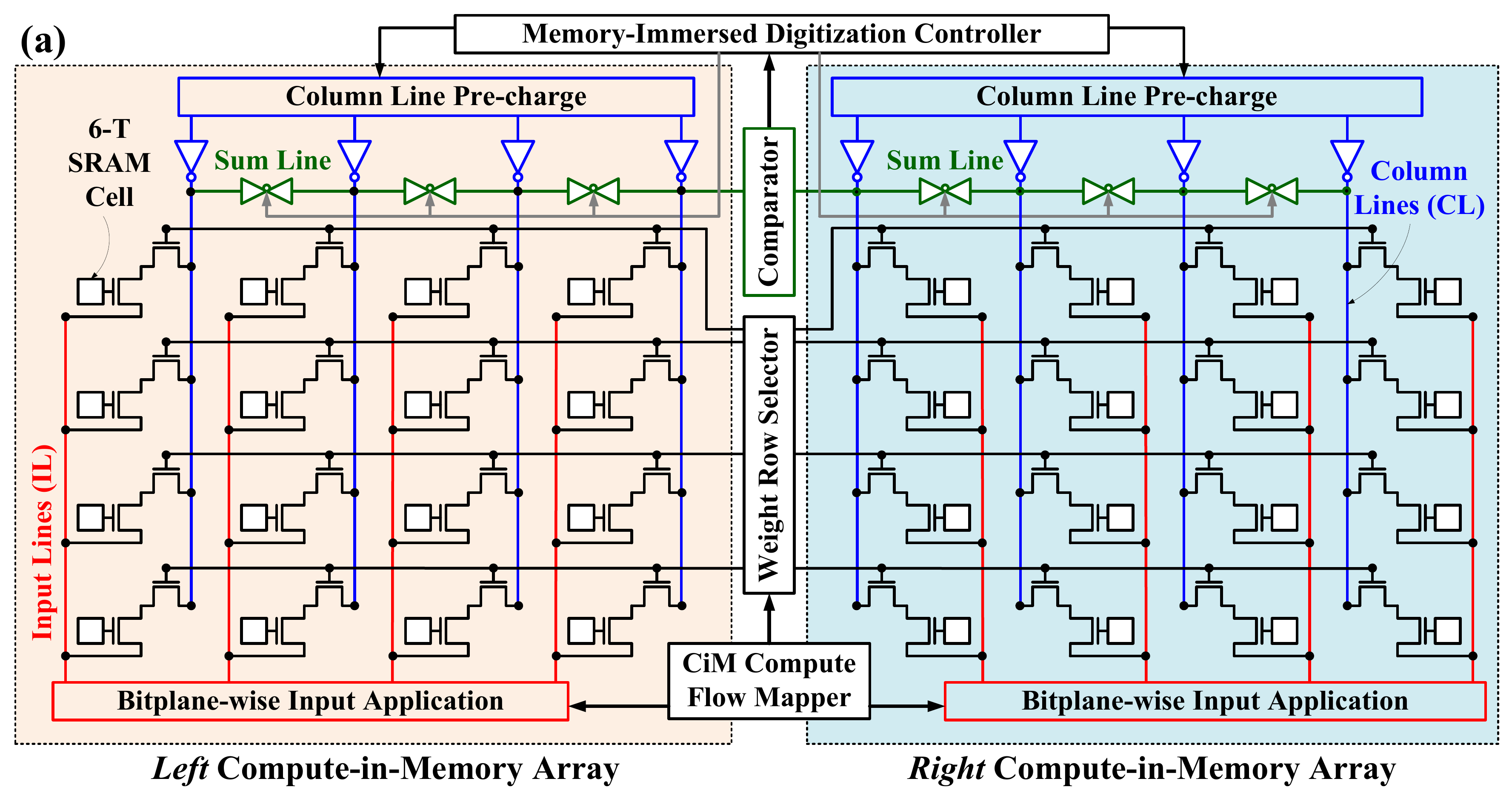}
  \includegraphics[width=0.4\linewidth]{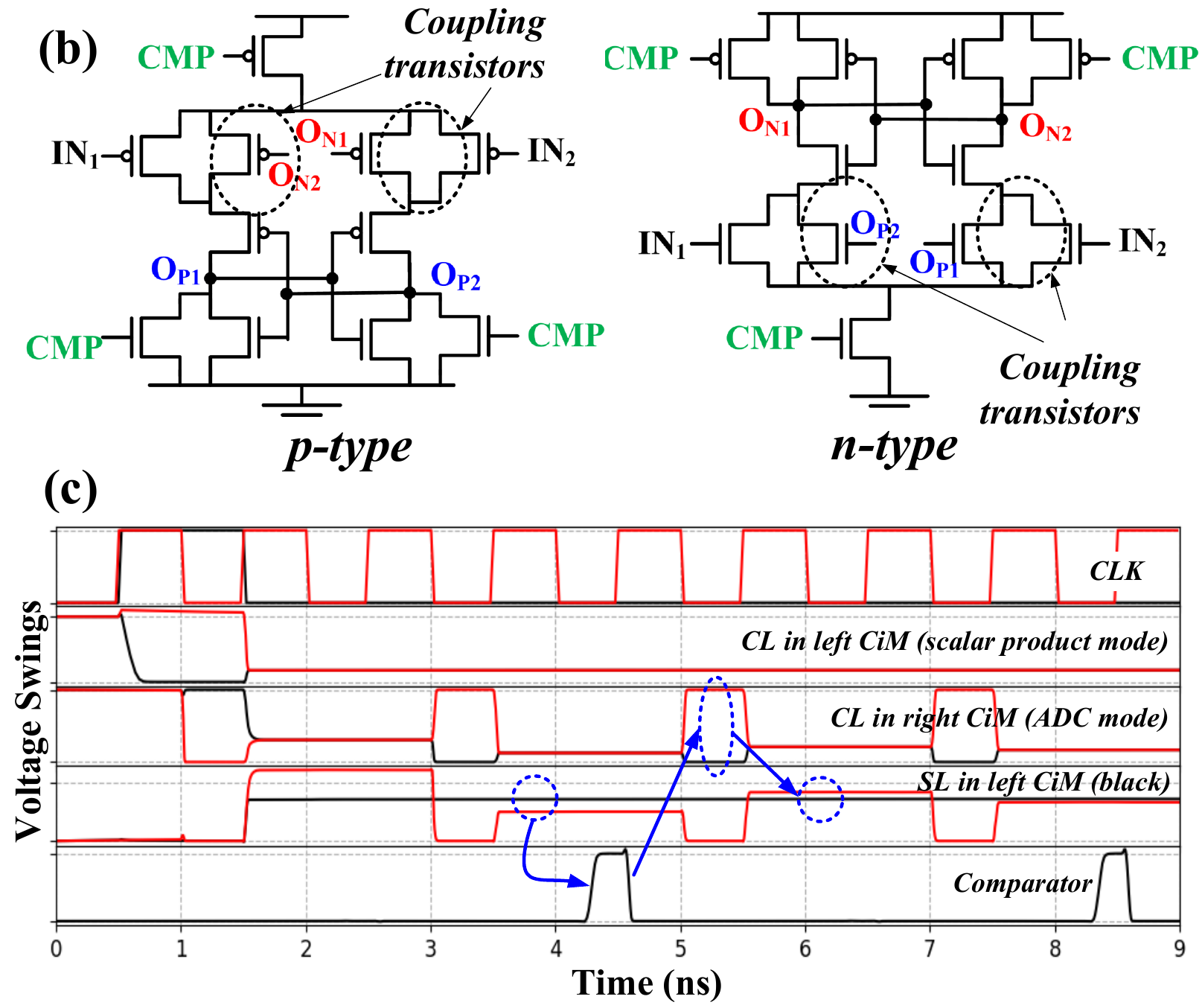}
  \caption{\textbf{Architecture and waveforms of SRAM-immersed ADC:} \textbf{(a)} Coupling of left-right memory arrays for memory-immersed digitization. When the left array computes within-memory scalar product, the right array digitizes analog-domain computed output. Both arrays switch their operation subsequently for collaborative digitization. \textbf{(b)} Clocked comparator design combining n-type and p-type counterparts for rail-to-rail voltage comparison. \textbf{(c)} Transient waveforms.} \vspace{-1pt}
  \label{fig:arch}
\end{figure*}

\begin{figure}[t!]
  \centering
  \includegraphics[width=0.8\linewidth]{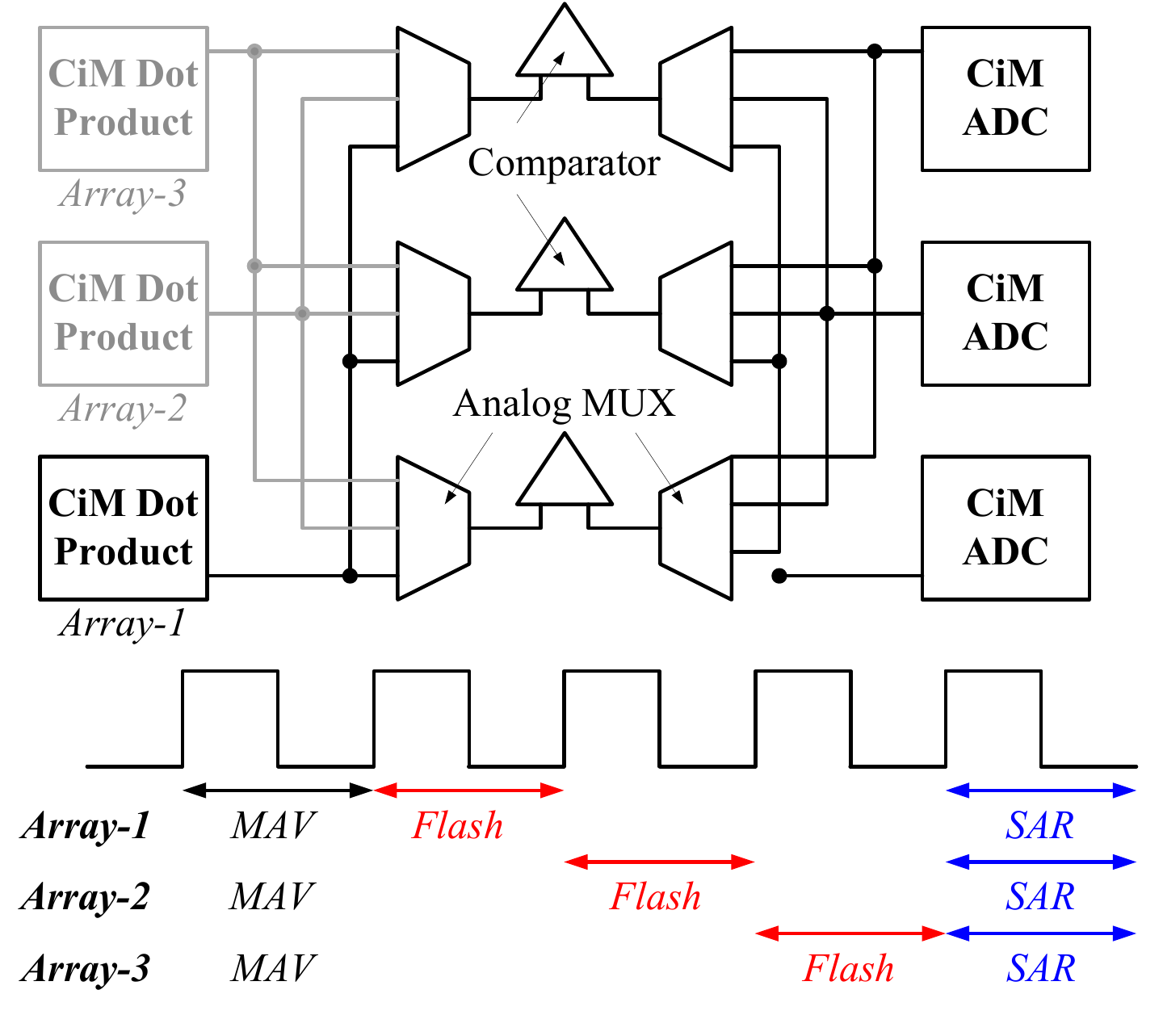}
  \caption{\textbf{Hybrid mode of SRAM-immersed ADC:} A dot product-configured CiM array is coupled to many ADC-configured arrays to the right for flash mode digitization of the initial most significant bits. After this, each left-array couples to the nearest right array to determine the remaining bits in SAR mode. Operational cycles are shown at the bottom.}
  \label{fig:chip}
\end{figure}

\section{Memory-Immersed Collaborative Analog-to-Digital Conversion}
\subsection{Coupling CiM arrays for collaborative digitization}
Fig. \ref{fig:arch} shows the realization of memory-immersed ADC. We specifically discuss our techniques and results for eight transistors (8T) compute-in-SRAM arrays which have been predominantly used in many platforms \cite{shukla2022mc,nasrin2021compute,nasrin2020supported,yu202016k,yu202265,shukla2020mc}. Unlike 6T compute-in-SRAM, 8T cells are less susceptible to bit disturbs due to decoupled write and inference ports and thus are more amenable to technology scaling. Nonetheless, our propositions on memory-immersed ADC apply to other memory types, including 10-T compute-in-SRAM/eDRAM \cite{raman2021compute,xie202116,ha202236,biswas2018conv} as well as non-volatile memory crossbars \cite{rahimifard2022higher,nasrin2019low}. 

In the proposed scheme, two proximal CiM arrays collaborate for within-memory digitization, as shown in Fig. 2(a). In the figure, when the left array computes the input-weight scalar product, the right array performs SRAM-immersed digitization on the generated analog-mode multiply-average (MAV) outputs. Both arrays switch their operating mode subsequently. Each array comprises 8-T cells, combining standard 6-T SRAM cells with a two-transistor weight-input product port shown in the figure. Memory cells for input-weight products are accessed using three control signals. Row lines (RL) are routed horizontally and select two-transistor weight-input product ports for operation. Input lines (IL) are routed vertically and apply input bits to the CiM array. Column lines (CL) are routed vertically and evaluate the 1-bit input-weight product. After the input-weight product completion on CL, the lines are merged on sum lines (SL) to compute the product sum in the charge domain. 

\begin{figure}[t!]
  \centering
  \includegraphics[width=\linewidth]{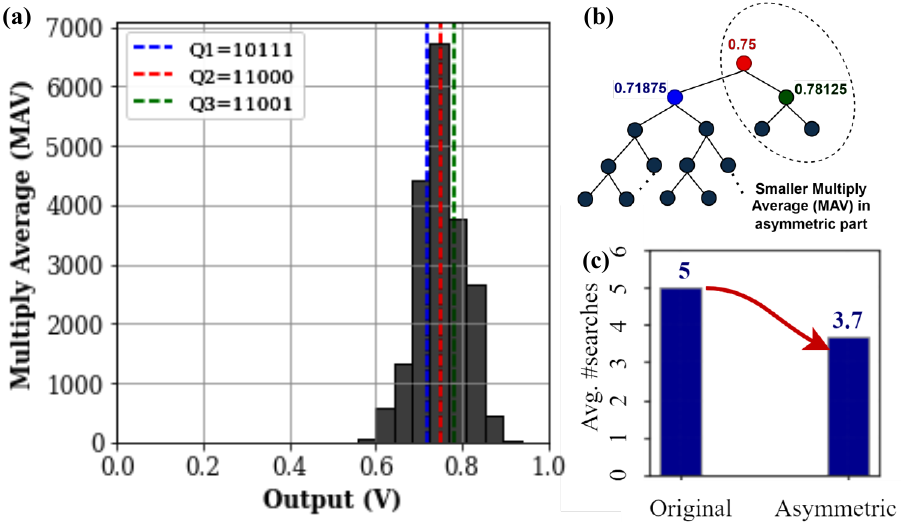}
  \caption{\textbf{Exploiting MAV statistics for ADC's time-efficiency:} \textbf{(a)} Distribution of MAV under the uniform distribution of input and weight bits for CiM scheme in Figure 2. \textbf{(b)} Asymmetric binary search for skewed MAV statistics. \textbf{(c)} For 5-bit data conversion, asymmetric search requires on average $\sim$3.7 comparisons, unlike symmetric binary search that requires $\sim$5 comparisons.}
  \label{fig:statistics}
\end{figure}

\begin{figure*}[t!]
  \centering
  \includegraphics[width=\linewidth]{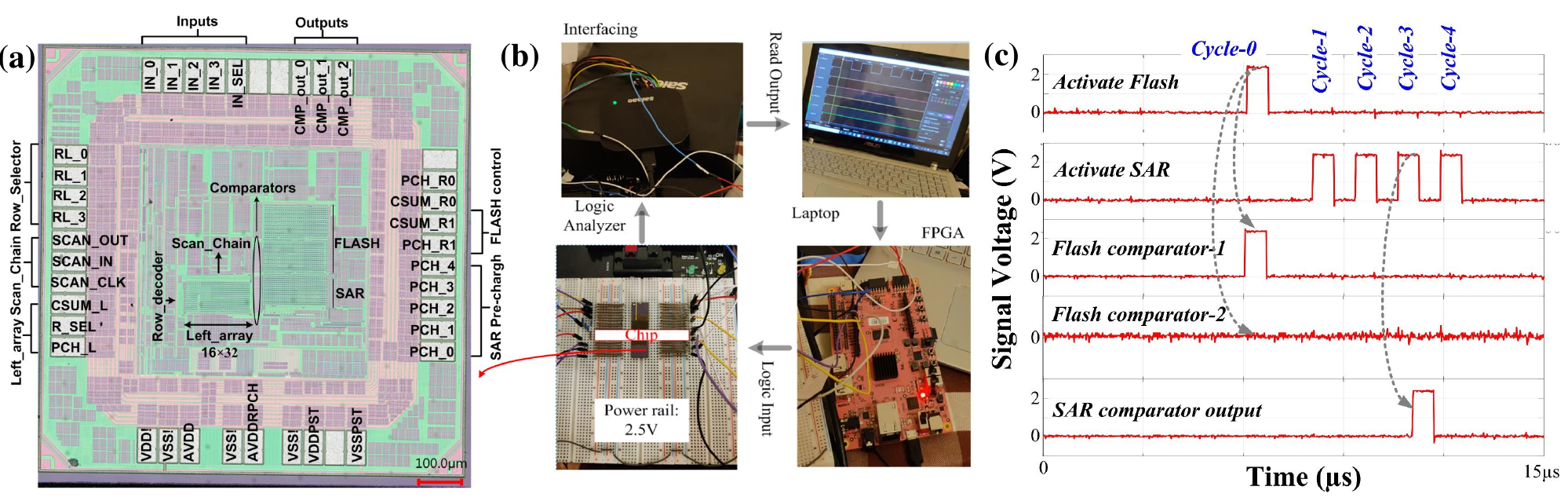}
  \caption{\textbf{Test-chip and measurements:} \textbf{(a)} Micrograph of fabricated design in 65 nm CMOS. Four compute-in-SRAM arrays, A1$-$A4 were fabricated. A1 interfaces with A2 to realize SRAM-immersed SAR ADC. A1 interfaces with A1$-$A4 to realize SRAM-immersed Flash ADC. \textbf{(b)} Measurement setup. \textbf{(c)} Measurement transient waveforms for hybrid SAR + Flash ADC operation.}
  \label{fig:chip}
\end{figure*}

For memory-immersed digitization of charge-domain product-sum computed in the left array, CLs in the right array realize the unit capacitors of a capacitive DAC formed within the memory array. A precharge transistor array is integrated with the column lines to generate the reference voltages, $V_\text{REF}$. In the successive approximation (SA) mode of digitization, the operation begins by precharging half of the column lines to supply voltage (VDD) and discharging the remaining to the ground (GND). The first reference voltage $V_\text{REF,0}$ is generated by summing the charges of all column lines using transmission gates in Fig. 2(a). The developed multiply average (MAV) voltage $V_\text{MAV}$ in the left CiM array is compared to $V_\text{REF,0}$ to determine the most significant bit $B_0$ of the digitized output. The next precharge state of memory-immersed capacitive DAC is determined, and the precharge and comparison cycles continue until $V_\text{MAV}$ has been digitized to necessary precision. Based on \cite{nasrin2021mf}, Fig. 2(b) shows the design of a comparator that couples n-type and p-type counterparts for rail-to-rail comparison. Fig. 2(c) shows the operational waveforms considering a DAC-free digital application of input vectors where we bit slice inputs to 1-bit, i.e., apply one input bit plane in one operating cycle. However, the operation of a memory-immersed digitizer is versatile and applicable to any compute-in-memory that computes product-sum output in the voltage/charge domain.    

Exploiting neighborhood CiM arrays for $V_\text{REF,0}$ generation for memory-immersed digitization provides many crucial advantages. \textit{First}, various non-idealities in analog-mode MAV computation such as the impact of the parasitic capacitance of column merge switches, wires, \textit{etc.}, become common-mode due to exploiting an identical array for $V_\text{REF,0}$ generation. Thus, the non-idealities only minimally impact the accuracy of digitization. \textit{Secondly}, collaborative digitization minimizes peripheral overheads. Only an analog comparator and simple modification in the precharge array are sufficient to realize a successive approximation search.  

Compared to traditional CiM approaches where a dedicated ADC is used at each array, interleaving of scalar product computation and digitization cycles in our scheme affects the achievable throughput. Meanwhile, with simplified low-area peripherals, more CiM arrays can be accommodated than prior works employing dedicated ADCs. Therefore, our scheme compensates for the overall throughput at the system level by operating many parallel CiM arrays. Especially the improved area efficiency of CiM arrays in our scheme minimizes the necessary exchanges from off-chip DRAMs to on-chip structures in mapping large DNN layers, a significant energy overhead in conventional techniques. 

\subsection{Hybrid SRAM-immersed Flash and SAR ADC Operation}
In addition to the nearest neighbor networking in Fig. 2, more intricate CiM networks can also be orchestrated for more time-efficient collaborative digitization in Flash and/or hybrid SAR + Flash mode. Fig. 3 shows an example networking scheme where \textit{Array-1} couples with three memory arrays to the right for collaborative digitization in Flash mode. Here, the three right arrays simultaneously generate the respective reference voltages $V_\text{REF,0}$--$V_\text{REF,2}$ for the Flash mode of digitization and to determine the first two most significant bits in one comparison cycle. Time steps for the hybrid mode digitization scheme are shown at the bottom of Fig. 3. CiM-configured memory arrays sequentially connect to all three ADC-configured arrays in parallel to digitize their first most significant bits, i.e., $D_0$--$D_1$. After this, each array couples to the nearest ADC-configured CiM array to the right for SAR mode digitization of the remaining bits in parallel across all arrays. In the discussed case, i.e., one-to-three coupling, hybridization reduces the energy for digitization since the initial references for the Flash mode are shared among CiM arrays and need not be generated individually, thus saving the reference generation energy. 

\subsection{Exploiting MAV Statistics for ADC's Time-Efficiency}
The above hybrid scheme for data conversion further benefits from exploiting the CiM-computed multiply average (MAV) statistics. In many CiM schemes, the computed MAV is not necessarily uniformly distributed. For example, in bit-plane-wise CiM processing, DACs are avoided by processing a one-bit input plane in one time step. Under single-ended processing using 8T cells, the column bit line discharges only when both the stored bit and applied input are `1.' Thus, even if weights and inputs are uniformly distributed, the probability of bit line discharge is 25\%, and MAV is skewed distributed, as shown in Fig. \ref{fig:statistics}(a). Similarly, in most DNNs, weights are regularized to be small, and activation layers such as rectified linear units (ReLU) are used to promote the sparsity of outputs. This, too, results in a skewed distribution of MAV. The skewed distribution of MAV can be leveraged by implementing an asymmetric binary search for digitization, as shown in Fig. \ref{fig:statistics}(b). In Fig. \ref{fig:statistics}(c), under the asymmetric search, the average number of comparisons reduces to $\sim$3.7 for 5-bit digitization compared to 5 comparisons under symmetric search, thus proportionally reducing the energy and latency for the operation. Even more, the proposed hybrid digitization scheme further exploits the asymmetric search since, in most cases, only two comparisons are sufficient (see right branches of the search tree in Fig. \ref{fig:statistics}(b)), which can be accelerated by Flash mode of digitization. 

\begin{figure*}[t!]
  \centering
  \includegraphics[width=0.95\linewidth]{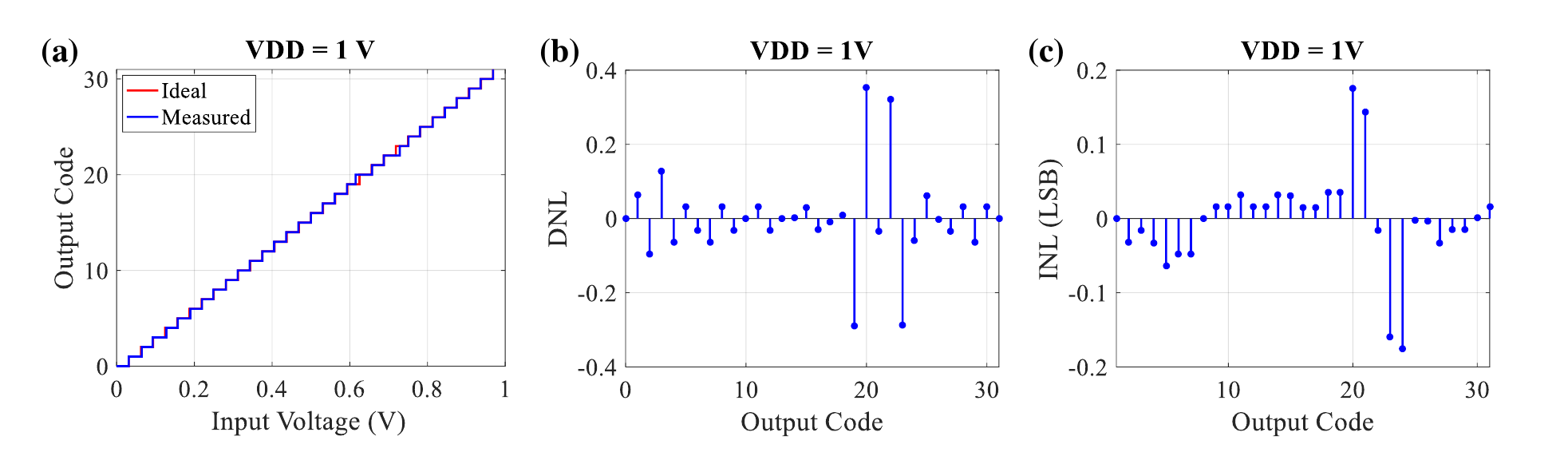}
  \caption{\textbf{Measured non-idealities of SRAM-immersed ADC:} (a) Output code vs. applied input voltage. (b) Differential and (c) integrated non-linearities.} \vspace{-10pt}
  \label{fig:INL_DNL}
\end{figure*}

\begin{figure}[t!]
  \centering
  \includegraphics[width=\linewidth]{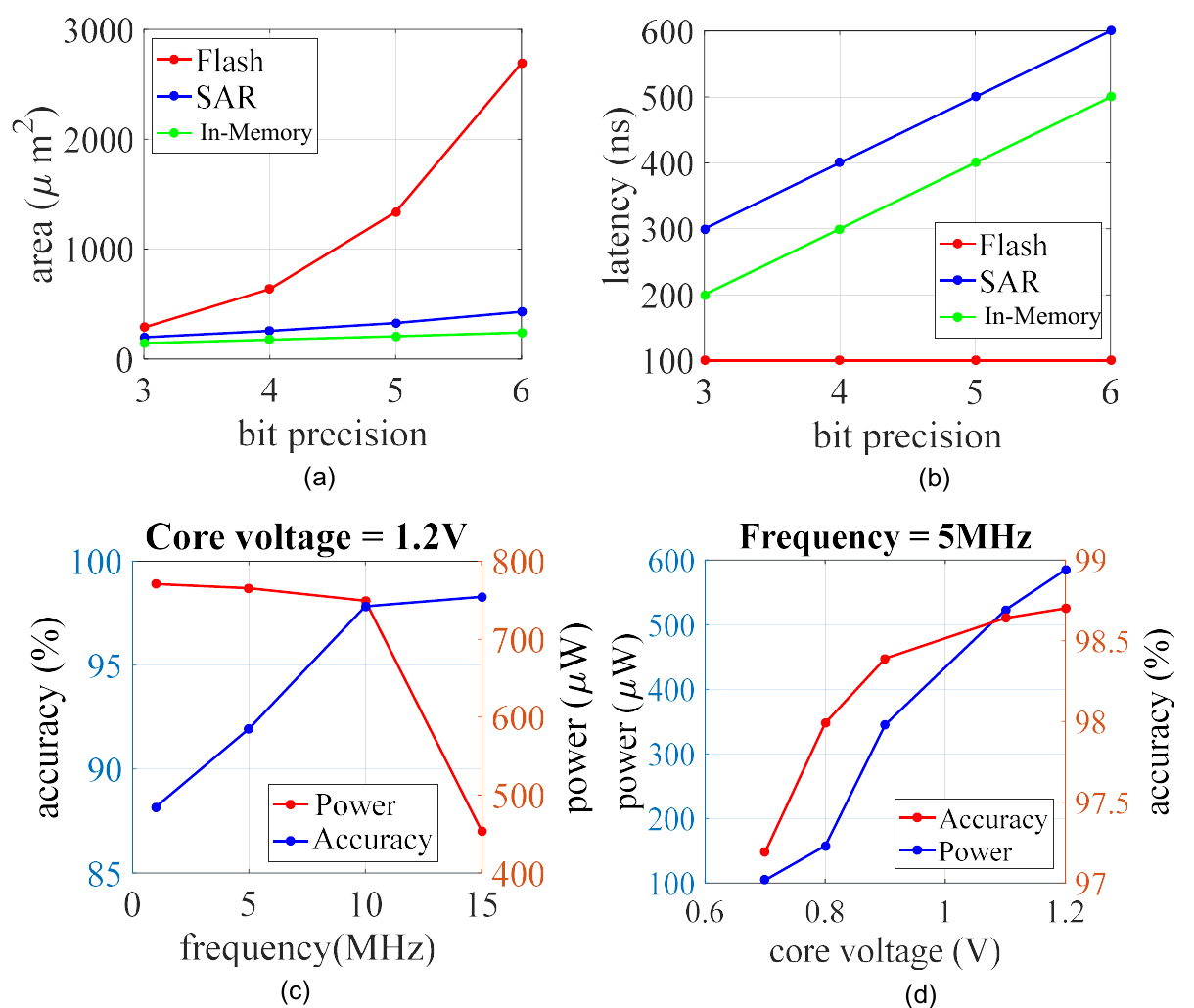}
  \caption{\textbf{Design space exploration of memory-immersed ADC:} Area in (a) and latency in (b) \textit{vs.} bit precision for different ADC styles. For memory-immersed ADC MNIST prediction accuracy and power \textit{vs.} frequency in (c) and \textit{vs.} operating voltage in (d). }
  \label{fig:design_space}
\end{figure}

\section{Test-Chip Design, Measurement Results, and Comparison to Traditional ADC}
A 65 nm CMOS test chip characterized the proposed SRAM-immersed ADC. Figs. \ref{fig:chip}(a, b) shows the fabricated chip's micrograph and measurement setup. Four compute-in-SRAM arrays A1$-$A4 of size 16$\times$32 were implemented. A1 couples with A2 to realize within-SRAM 5-bit SAR ADC in the designed chip. A1 couples with A2$-$A4 for within-SRAM 2-bit Flash ADC using the schemes discussed in Fig. 3. The coupling of CiM arrays can also be programmed to realize hybrid Flash-SAR ADC operations, such as obtaining the two most significant bits in Flash mode and the remaining in SAR.

Fig. \ref{fig:chip}(c) shows the transient waveforms of different control signals and comparator outputs, showing a hybrid Flash + SAR ADC operation. Flash mode is activated in the first comparison cycle where CiM arrays A2$-$A4 generates the corresponding reference voltages, and the first two bits of MAV digitization are extracted. Subsequently, the operation switches to SAR mode, where the remaining digitization bits are obtained by engaging A2 alone with A1. In the last four cycles, A3 \& A4 become free to similarly operate on a proximal CiM array to digitize MAV in SAR mode. Fig. \ref{fig:INL_DNL}(a) shows the measured staircase plot of input voltage to output codes and the comparison to an ideal staircase, showing near-ideal characteristics. The corresponding differential and integrated non-linearities at various output codes are shown in Figs. \ref{fig:INL_DNL}(b, c) which are always below 0.5 bits. 


\begin{center}
\textbf{Table I: Comparison of 5-bit in-memory ADC with 10 MHz clock against SAR and Flash architectures} \\ 
\vspace{5pt}
\begin{tabular}{ p{3cm} p{1.3cm} p{1.3cm} p{1.3cm}} 
\hline
\textbf{Architecture }& Tech. & Area ($\mu$m$^2$) & Energy (pJ) \\ \hline \hline
\textbf{SAR} \cite{jiang2021analog} & 40 nm & 5235.20 & 105 \\ \hline
\textbf{Flash} \cite{jiang2021analog} & 40 nm & 10703.36 & 952 \\ \hline
\textbf{In-Memory} (ours) & 65 nm & 207.8 & 74.23 \\ \hline
\end{tabular}
\end{center}

Fig. \ref{fig:design_space} and Table I show the design space exploration of proposed memory-immersed ADC compared to other ADC styles. In Fig. \ref{fig:design_space}(a), leveraging in-memory structures for capacitive DAC formation, the proposed in-memory ADC is more area efficient than Flash and SAR styles. Significantly, Flash ADC's size increases exponentially with increasing bit precision. In Fig. \ref{fig:design_space}(b), SAR ADC's latency increases with bit precision while Flash ADC can maintain a consistent latency but at the cost of the increasing area as shown in Fig. \ref{fig:design_space}(a). A hybrid data conversion in the proposed in-memory provides a middle ground, i.e., lower latency than in SAR ADC. Figs. \ref{fig:design_space}(c-d) show the impact of supply voltage and frequency scaling on in-memory ADC's power and accuracy for MNIST character recognition. At increasing frequency 

\section{Conclusions}
We have presented a memory-immersed ADC that obviates the area overheads of a traditional CiM-based deep learning inference design. Exploiting the \textit{multi-functionality} of CiM arrays, the silicon space can be maximally exploited for CiM-based processing to support a higher degree of parallelism and to minimize exchanges with off-chip memories in footprint-constrained designs. In the proposed scheme, proximal CiM arrays collaborate to efficiently realize various data conversion schemes such as Flash, successive approximation (SA), and their hybrid steps. The results were demonstrated using 65 nm CMOS technology demonstrating impressive energy and area-efficiency advantages over the traditional designs. 

\vspace{10pt}
\noindent \textbf{Acknowledgement:} This work was supported in part by COGNISENSE, one of seven centers in JUMP 2.0, a Semiconductor Research Corporation (SRC) program sponsored by DARPA, and NSF CAREER Award \#2046435.

\bibliographystyle{IEEEtran}
\bibliography{main.bib}
\vfill

\end{document}